\documentclass[12pt]{article}
\usepackage{graphicx}

\newcommand\pubdate{\today}

\newcommand\pubnumber{FERMILAB-CONF-11-536-T}

\def\Title#1{\begin{center} {\Large #1 } \end{center}}
\def\Author#1{\begin{center}{ \sc #1} \end{center}}
\def\Address#1{\begin{center}{ \it #1} \end{center}}

\newcommand\pubblock{\rightline{\begin{tabular}{l} \pubnumber\\
         \pubdate  \end{tabular}}}
\newenvironment{Abstract}{\begin{center}{\bf Abstract}\end{center} \bigskip \begin{quotation}  }{\end{quotation}}
\newenvironment{Presented}{\begin{quotation} \begin{center} 
             PRESENTED AT\end{center}\bigskip 
      \begin{center}\begin{large}}{\end{large}\end{center} \end{quotation}}
\def\Acknowledgements{\bigskip  \bigskip \begin{center} \begin{large}
             \bf ACKNOWLEDGEMENTS \end{large}\end{center}}

\def\npb#1#2#3{{\it Nucl.\ Phys.} {\bf B#1}, #2 (#3)}
\def\plb#1#2#3{{\it Phys.\ Lett.} {\bf B\,#1}, #2 (#3)}
\def\prd#1#2#3{{\it Phys.\ Rev.} {\bf D#1}, #2 (#3)}

\def\U4s{\Upsilon (4s)}
\def\oB0{\overline{B^0} }

\def\ra{\rightarrow}

\newcommand{\Eq}[1]{Eq.~(\ref{eq#1})}

\def\beq{\begin{equation}}
\def\eeq#1{\label{#1}\end{equation}}
\def\eeqn{\end{equation}}

\def\beqa{\begin{eqnarray}}
\def\eeqa#1{\label{#1}\end{eqnarray}}
\def\eeqan{\end{eqnarray}}

\let\bar=\overbar

\def\etal{{\it et al.}}

\def\Dslash{\not{\hbox{\kern-4pt $D$}}}
\def\dslash{\not{\hbox{\kern-2pt $\del$}}}

\def\msb{{\bar{\ssstyle M \kern -1pt S}}}

\begin{document}
\begin{titlepage}
\pubblock

\vfill

\Title{B-Meson and Neutrino Oscillation: A Unified Treatment}
\vfill
\Author{Boris Kayser \footnote[1
]{E-mail: boris@fnal.gov} }
\Address{Theoretical Physics Department, Fermilab, P.O. Box 500, Batavia, IL 60510  USA}
\vfill

\begin{Abstract}
We present a unified treatment of the quantum mechanics of $B$-factory and neutrino oscillation experiments. While our approach obtains the usual phenomenological predictions for these experiments, it does so without having to invoke perplexing Einstein-Podolsky-Rosen correlations or non-intuitive kinematical assumptions.
\end{Abstract}
\vfill

\begin{Presented}
The Ninth International Conference on\\
Flavor Physics and CP Violation\\
(FPCP 2011)\\
Maale Hachamisha, Israel,  May 23--27, 2011
\end{Presented}
\vfill

\end{titlepage}
\setcounter{footnote}{0}

\section{Introduction} \label{intro}

The quantum mechanics of mixing during propagation is at the heart of both $B$-factory and neutrino oscillation experiments. In this paper, we will treat both these experiments in the same way. Our treatment has several advantages. In dealing with the $B$-factory experiments, it avoids having to invoke real but nonetheless puzzling Einstein-Podolsky-Rosen correlations. In dealing with neutrino oscillation, our approach avoids the non-intuitive assumption that all the interfering neutrino mass eigenstates in a beam have the same energy.

\section{$B$-Factory Experiments} \label{sec2}

\begin{figure}[htb]
\begin{center}
\includegraphics[width=0.8\textwidth]{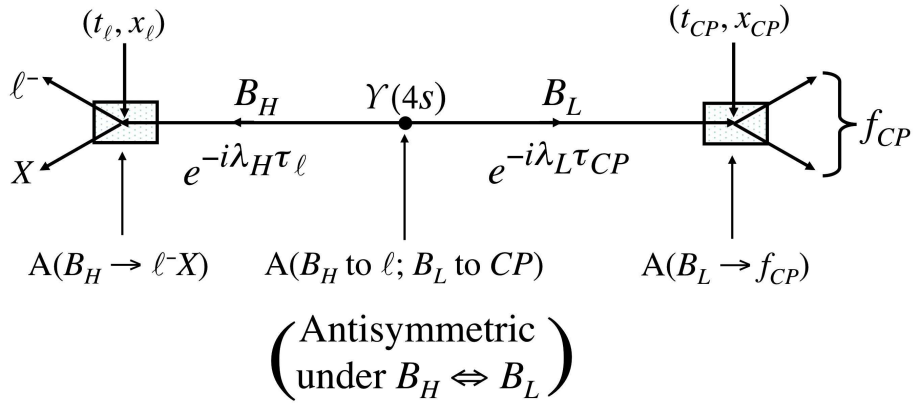}
\caption{One of two coherent contributions to the process in which an $\U4s$ decays into a neutral $B$ pair, after which one $B$ decays semileptonically to a negative lepton $\ell^-$ and other particles $X$, and the other $B$ decays to a hadronic CP eigenstate $f_{CP}$. The various features and factors appearing in the figure are defined in the text.}
\label{f1}
\end{center}
\end{figure}

In a typical $B$-factory experiment studying CP violation, an electron and positron collide and form an $\U4s$. The $\U4s$, a $b \bar{b}$ bound state with intrinsic spin $S=1$, then decays to a pair of $B$ mesons. We are interested in the case where these $B$ mesons are a $B^0$ and a $\oB0$, neutral particles that then undergo $B^0 - \oB0$ mixing. Viewing the process in the $\U4s$ rest frame, as in Fig.~\ref{f1}, 
we are especially interested in events in which one of the $B$ mesons decays semileptonically into a negatively charged lepton $\ell^-$ plus other particles $X$ at a spacetime point $(t_\ell, x_\ell)$ in the $\U4s$ rest frame, while the other $B$ meson decays hadronically into a CP eigenstate $f_{CP}$ at some other spacetime point $(t_{CP}, x_{CP})$ in the $\U4s$ rest frame. We shall calculate the amplitude for the entire sequence, from the $\U4s$ decay through the $B$ decays of interest \cite{ref1}. In doing so, we shall make use of the fact that if an unstable particle has mass $m$ and width $\Gamma$, adding up to a complex mass $\lambda = m-i \Gamma / 2$, then the amplitude for this particle to propagate for a proper time $\tau$ in its own rest frame is $\exp (-i \lambda \tau)$ \cite{ref1}. 
(If this propagation is through a distance $x$ during a time $t$ in some frame in which the particle has momentum $p$ and energy $E$, then the non-decaying part of $\exp (-i \lambda \tau)$, namely $\exp (-i m \tau)$, is simply the familiar quantum-mechanical plane-wave factor $\exp [i(px - Et)]$.)

In order to employ the simple propagation amplitude $\exp (-i \lambda \tau)$, we work in the $B$ mass eigenstate basis. There are two neutral $B$ mass eigenstates: the heavier one $B_H$, and the lighter one $B_L$. Since these $B$ mesons are spinless, while the $\U4s$ has $S=1$, the decay $\U4s \ra BB$ leaves the $B$ mesons in a $p$ wave. Since one cannot have two identical spinless bosons in a $p$ wave, one of the daughter $B$ mesons must be a $B_H$, while the other is a $B_L$. 
There are then two scenarios that contribute to the sequence $\U4s \ra BB \ra (\ell^- X)(f_{CP})$. In the first, pictured  in Fig.~\ref{f1}, it is a $B_H$ that heads for the point $(t_\ell, x_\ell)$ and there decays into $\ell^- X$, and a $B_L$ that heads for the point $(t_{CP}, x_{CP})$ and there decays into $f_{CP}$. 
In the second scenario, it is a $B_L$ that travels to $(t_\ell, x_\ell)$ and decays into $\ell^- X$, and a $B_H$ that travels to $(t_{CP}, x_{CP})$ and decays into $f_{CP}$. Since the mass difference $m_{B_H} - m_{B_L} = 3 \times 10^{-4}$ eV is exceedingly tiny, one will not know which scenario was responsible for any given event. Therefore, the amplitudes for the two scenarios must be added coherently.

For the scenario in which $B_H$ yields $\ell^- X$ and $B_L$ yields $f_{CP}$, the amplitude is the product of factors shown in Fig.~\ref{f1}. The first of these factors, $A(B_H \;\mathrm{to} \;\ell ; B_L \;\mathrm{to \; CP})$, is the amplitude for $\U4s$ to decay into a $B_H$ headed towards the point $(t_\ell, x_\ell)$ and a $B_L$ headed towards $(t_{CP}, x_{CP})$. The only feature of this amplitude that is relevant for us is that, owing to the antisymmetry of the $p$ wave, it is antisymmetric under $B_H \Leftrightarrow B_L$. 
Next, there is the amplitude $\exp (-i \lambda_H \tau_\ell)$ for the $B_H$, with complex mass $\lambda_H$, to travel from the $\U4s$ decay point, which we shall call $(0,0)$, to the point $(t_\ell, x_\ell)$. The quantity $\tau_\ell$ is the proper time that elapses in the $B_H$ rest frame during this journey. There is also the analogous amplitude $\exp (-i \lambda_L \tau_{CP})$ for the $B_L$, with complex mass $\lambda_L$, to travel from $(0,0)$ to $(t_{CP}, x_{CP})$. The quantity $\tau_{CP}$ is the proper time that elapses in the $B_L$ rest frame during this trip. Finally, there are the amplitudes $A(B_H \ra \ell^- X)$ and $A(B_L \ra f_{CP})$ for the $B_H$ to decay to $\ell^- X$ and the $B_L$ to $f_{CP}$. Thus, including the scenario not pictured in Fig.~\ref{f1}, the amplitude {\it Amp} for $\U4s \ra BB$, following which one $B \ra \ell^- X$ after traveling for a proper time $\tau_\ell$, and the other $B \ra f_{CP}$ after traveling for a proper time $\tau_{CP}$, is given by
\beq
\mathrm{{\it Amp}} \propto e^{-i \lambda_H \tau_\ell}\, e^{-i \lambda_L \tau_{CP}} A(B_H \ra \ell^- X) \,A(B_L \ra f_{CP}) - (B_H \Leftrightarrow B_L) ~~.
\label{eq1}
\eeqn
We note that this {\it Amp} is Lorentz invariant.

Using the fact that $B_H$ and $B_L$ have essentially the same width $\Gamma$, we may write
\beq
\lambda_{H,\,L} = m \pm \frac{\Delta m}{2} - i \frac{\Gamma}{2} ~~,
\label{eq2}
\eeqn
where $m \equiv (m_{B_H} + m_{B_L}) / 2$, and $\Delta m \equiv (m_{B_H} - m_{B_L})$. For the $B_{H,\,L}$ decay amplitudes, we use the standard relation
\beq
B_{H,\,L} = \frac{1}{\sqrt{2}} \left[ B^0 \pm e^{-2i \delta_{CKM}^\mathrm{mix}} \; \oB0 \right] ~~,
\label{eq3}
\eeqn
in which $\delta_{CKM}^\mathrm{mix}$ is a CP-violating $B^0 - \oB0$ mixing phase coming from the CKM quark mixing matrix. Since only a $\oB0$, but not a $B^0$, can decay semileptonically to a negatively-charged lepton $\ell^-$, this relation immediately yields the relevant factors in $A(B_{H,\,L} \ra \ell^- X)$. For the decays to $f_{CP}$, we assume that $B^0 \ra f_{CP}$ is dominated by a single diagram, so that we may write the $B^0$ and $\oB0$ decay amplitudes as
\beq
A(B^0 \ra f_{CP}) = M e^{i \delta_{CKM}^f}\, e^{i \alpha_{ST}}
\label{eq4}
\eeqn
and
\beq
A(\oB0 \ra f_{CP}) = \eta_f M  e^{-i \delta_{CKM}^f}\, e^{i \alpha_{ST}}~~.
\label{eq5}
\eeqn
Here, $M$ is the magnitude of the dominating diagram, $\delta_{CKM}^f$ is its CP-violating CKM phase, $\alpha_{ST}$ is its CP-even strong interaction phase, and $\eta_f$ is the CP parity of the CP eigenstate $f_{CP}$.

From Eqs.~(\ref{eq1}) - (\ref{eq5}), we find that the probability $P[\Upsilon \ra BB \ra (\ell^- X)(f_{CP})]$ that $\U4s \ra BB$, and then one $B \ra \ell^- X$ after $\tau_\ell$ and the other $B \ra f_{CP}$ after $\tau_{CP}$, is given by
\beq
P[\Upsilon \ra BB \ra (\ell^- X)(f_{CP})] \propto |\mathrm{{\it Amp}}|^2 \propto e^{-\Gamma(\tau_{CP} + \tau_\ell)} \{1 - \eta_f \sin \phi \sin [\Delta m (\tau_{CP} - \tau_\ell)] \} ~~.
 \label{eq6}
 \eeqn
Here, $\phi$ is the CP-violating phase defined by
 \beq
 \phi \equiv 2 ( \delta_{CKM}^\mathrm{mix} +  \delta_{CKM}^f) ~~.
 \label{eq7}
 \eeqn
 The probability of \Eq{6} is the usual result \cite{ref2}, except that times in the $\U4s$ rest frame are replaced by proper times in the $B$ rest frames. This is a negligible correction. Much more importantly, we have derived the usual result without invoking any puzzling Einstein-Podolsky-Rosen correlations. In the usual treatment, one imagines to begin with that the decay $B \ra \ell^- X$, at a time $t_\ell$ in the $\U4s$ rest frame, is the first of the two $B$ decays to occur. Then, since only a $\oB0$, but not a $B^0$, can undergo the decay $B \ra \ell^- X$, and since one cannot have two identical spinless bosons in a $p$ wave, one concludes that at the time $t_\ell$ in the $\U4s$ rest frame, the surviving $B$ must be a pure $B^0$. 
 This surviving $B$ then undergoes $B^0 - \oB0$ mixing, and evolves according to the Schr\"{o}dinger time evolution equation, until it decays into $f_{CP}$. This is a perfectly legitimate approach, but it does leave one wondering how the surviving $B$ ``knows'' how and when the other $B$, with which it is not communicating, decayed. Our approach avoids this puzzle.

\section{Neutrino Oscillation} \label{sec3}

\begin{figure}[htb]
\begin{center}
\includegraphics[width=0.8\textwidth]{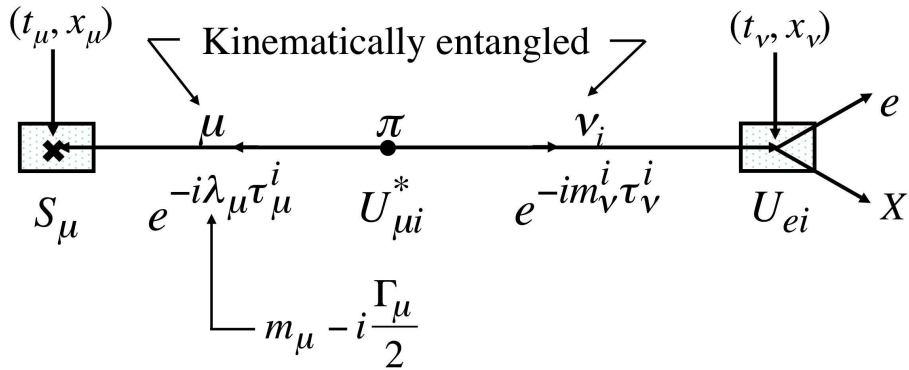}
\caption{The contribution of neutrino mass eigenstate $\nu_i$ to the process in which a pion undergoes the decay $\pi \ra \mu + \nu$, after which the muon interacts with matter at some spacetime point, and the neutrino interacts in a detector at some other spacetime point, producing an electron $e$ and other particles $X$. The various features and factors appearing in the figure are defined in the text.}
\label{f2}
\end{center}
\end{figure}

Now let us treat neutrino oscillation in the same way as we have treated $B$-factory experiments. Consider the illustrative process pictured in Fig.~\ref{f2}.
There, a neutrino is produced via the two-body decay $\pi \ra \mu + \nu$. We shall describe the process in the pion rest frame. The muon travels from the spacetime point where the pion decays, which we shall call $(0,0)$, to the spacetime point $(t_\mu, x_\mu)$, where it interacts with the matter surrounding the pion decay region. The neutrino journeys from $(0,0)$ to the spacetime point $(t_\nu, x_\nu)$, where it interacts in a detector and creates an electron. The propagating neutrino is one or another of the neutrino mass eigenstates $\nu_i$.

As we did for the $B$-factory experiments, we shall calculate the amplitude for the entire sequence shown in Fig.~\ref{f2}. For the decay $\pi \ra \mu + \nu_i$, the relevant factor in the amplitude is $U^*_{\mu i}$, where $U$ is the Pontecorvo-Maki-Nakagawa-Sakata leptonic mixing matrix---the leptonic analogue of the CKM quark mixing matrix. Similarly, for the creation of an electron in the detector by $\nu_i$, the relevant factor in the amplitude is $U_{ei}$. For the interaction of the muon in matter, we shall call the amplitude $S_\mu$. 
For the propagation of the neutrino $\nu_i$, the amplitude is $\exp (-i m_\nu^i \tau_\nu^i )$, as explained when we discussed $B$ meson propagation. Here, $m_\nu^i$ is the mass of $\nu_i$, and $\tau_\nu^i$ is the proper time that elapses in the $\nu_i$ rest frame during its journey from $(0,0)$ to $(t_\nu, x_\nu)$. In the phase $m_\nu^i \tau_\nu^i$, we are neglecting the extremely tiny neutrino decay width.  
However, we are allowing, through the index $i$ on $\tau_\nu^i$, for the fact that the proper time that elapses in the $\nu_i$ rest frame while the $\nu_i$ travels from $(0,0)$ to the given interaction point $(t_\nu, x_\nu)$ depends on the $\nu_i$ energy in the pion rest frame, hence on the $\nu_i$ mass, and consequently on which $\nu_i$ is involved. [In the $B$-factory experiments, the analogous effect---the dependence of $\tau_\ell$ or $\tau_{CP}$ on whether it is a $B_H$ or a $B_L$ that is propagating---is negligible. 
For example, in the $B_H$ propagation amplitude $\exp (-i \lambda_H \tau_\ell)$, the factor $\lambda_H = m + \Delta m /2 -i \Gamma /2$ certainly involves a term of 1st order in  $\Delta m \equiv m_{B_H} - m_{B_L}$, but the dependence of the proper time $\tau_\ell$ on $\Delta m$ is only 2nd order in $(m_{B_H} - m_{B_L}) / (m_{B_H} + m_{B_L}) \simeq 3 \times 10^{-14}$.] 
Finally, for the propagation of the muon, the amplitude is $\exp (-i \lambda_\mu \tau_\mu ^i)$, where $\lambda_\mu = m_\mu  -i \Gamma_\mu /2$ includes both the muon mass $m_\mu$ and its width $\Gamma_\mu$, and $\tau_\mu ^i$ is the proper time that elapses in the muon rest frame during its journey from $(0,0)$ to the point $(t_\mu, x_\mu)$ where it interacts. By giving $\tau_\mu ^i$ an index $i$, we are taking into account the fact that the muon and the neutrino are kinematically entangled. The energies of both of these particles in the pion rest frame depend on the mass of the emitted neutrino, so that they depend on which $\nu_i$ it is. Consequently, the proper times that elapse in the rest frames of the muon and the neutrino both depend on which $\nu_i$ is emitted, at least in principle.

Since we cannot make measurements that determine which $\nu_i$ was involved in any given event without destroying the oscillation pattern, the amplitudes for the contributions of the different $\nu_i$ must be added coherently. Thus, including all the factors shown in Fig.~\ref{f2}, the amplitude {\it Amp} for the entire sequence of events pictured there is given by
\beq
\mathrm{{\it Amp}} = \sum_{i=1,2,3} S_\mu\, e^{{\textstyle -i(m_\mu - i \Gamma_\mu /2) \tau_\mu^i}} \, U_{\mu i}^* \, e^{{\textstyle -i m_\nu^i \tau_\nu^i}}  \, U_{ei} ~~.
\label{eq8}
\eeqn
Here, we assume for the sake of illustration that there are just three neutrino mass eigenstates $\nu_i$. We note that the amplitude of \Eq{8} is Lorentz invariant.

How do the muon and neutrino propagation amplitudes $\exp [-i(m_\mu - i \Gamma_\mu /2)\tau_\mu^i]$ and $\exp (-i m_\nu^i \tau_\nu^i)$ actually depend on the index $i$? In the propagation amplitude for the muon,
\beq
\tau_\mu^i = \frac{1}{m_\mu} (E_\mu^i t_\mu - p_\mu^i x_\mu) ~~,
\label{eq9}
\eeqn
where $E_\mu^i$ and $p_\mu^i$ are, respectively, the muon energy and momentum in the pion rest frame when the emitted neutrino is $\nu_i$. In evaluating this expression, we choose a muon interaction spacetime point $(t_\mu,\, x_\mu)$ at which $x_\mu$ and $t_\mu$ are related by 
\beq
x_\mu = v_\mu^0 t_\mu = \frac{p_\mu^0}{E_\mu^0} t_\mu ~~.
\label{eq10}
\eeqn
Here, $v_\mu^0,\; p_\mu^0$, and $E_\mu^0$ are, respectively, the velocity, momentum, and energy the muon would have in the pion rest frame if all neutrinos were massless. The spacetime points passed by the peak of the wave packet describing the muon propagation in greater detail than is needed here would satisfy \Eq{10} to an excellent approximation.

From \Eq{9}, the difference between the muon travel proper times corresponding to $\pi \ra \mu + \nu_i$ and $\pi \ra \mu + \nu_j$ is given by 
\beq
\tau_\mu^i - \tau_\mu^j = \frac{1}{m_\mu} [(E_\mu^i - E_\mu^j) t_\mu - (p_\mu^i -  p_\mu^j)x_\mu] ~~.
\label{eq11}
\eeqn
Now, for given neutrino mass $m_\nu$, the energy $E_\mu$ of the recoiling muon in the pion rest frame is given by
\beq
E_\mu = \frac{m_\pi^2 + m_\mu^2 - m_\nu^2}{2m_\pi} ~~,
\label{eq12}
\eeqn
where $m_\pi$ is the pion mass. Thus, we have
\beq
E_\mu^i - E_\mu^j = -\frac{1}{2m_\pi} \Delta m^2_{ij} ~~,
\label{eq13}
\eeqn
where $\Delta m^2_{ij} \equiv (m_\nu^i)^2 - (m_\nu^j)^2$. Furthermore, for given $E_\mu,\; p^2_\mu = E^2_\mu - m^2_\mu$. so that
\beq
\frac{dp_\mu}{d(m^2_\nu)} = \frac{E_\mu}{p_\mu} \frac{dE_\mu}{d(m^2_\nu)}  ~~.
\label{eq14}
\eeqn
Thus, to lowest order in the squares of the neutrino masses,
\beq
p_\mu^i - p_\mu^j = \frac{E_\mu^0}{p_\mu^0} \left( - \frac{1}{2m_\pi}\right)  \Delta m^2_{ij} ~~.
\label{eq15}
\eeqn
From Eqs. (\ref{eq11}),  (\ref{eq10}), (\ref{eq13}), and  (\ref{eq15}), we then have, to lowest (i.e., first) order in the squares of the neutrino masses,
\beq
\tau_\mu^i - \tau_\mu^j = \frac{t_\mu}{m_\mu} \left[ -\frac{\Delta m^2_{ij}}{2m_\pi}\right] \left[ 1 -  \frac{E_\mu^0}{p_\mu^0}\frac{p_\mu^0}{E_\mu^0} \right] = 0 ~~.
\label{eq16}
\eeqn
That is, to lowest order, the muon propagation amplitude, $\exp (-im_\mu \tau_\mu^i) \exp [ -(\Gamma_\mu /2) \tau_\mu^i]$, actually does not depend on $i$. This was first noticed by Akhmedov and Smirnov \cite{ref3}. The factor $\exp (-im_\mu \tau_\mu^i)$ will have no effect on the absolute square of the amplitude of \Eq{8} for the sequence of events in Fig.~\ref{f2}. The factor $\exp [ -(\Gamma_\mu /2) \tau_\mu^i]$, which decays with time, will influence the dependence of the event rate for the sequence on how long the muon travels before it interacts, but it will not affect the neutrino oscillation pattern. Since it is only the latter in which we are ultimately interested, we can drop the entire muon propagation amplitude from the amplitude of \Eq{8}.

In the propagation amplitude for the neutrino $\nu_i,\; \exp (-i m_\nu^i \tau_\nu^i)$, we have
\beq
m_\nu^i \tau_\nu^i  = E_\nu^i t_\nu -p_\nu^i x_\nu ~~,
\label{eq17}
\eeqn
where $E_\nu^i$ and $p_\nu^i$ are, respectively, the $\nu_i$ energy and momentum in the pion rest frame.  Since in practice neutrinos are ultra-relativistic, we choose a neutrino interaction spacetime point $(t_\nu,\; x_\nu)$ at which $t_\nu = x_\nu \equiv L^0$. From the relation $E_\nu^i = [m_\pi^2 + (m_\nu^i)^2 - m_\mu^2] / 2m_\pi$, we find that the energies of two different neutrino mass eigenstates $\nu_i$ and $\nu_j$ differ by
\beq
E_\nu^i - E_\nu^j = \frac{1}{2m_\pi} \Delta m^2_{ij} ~~.
\label{eq18}
\eeqn
From the relation $(p_\nu^i)^2 = (E_\nu^i)^2 - (m_\nu^i)^2$, we find that, to lowest order in $\Delta m^2_{ij}$, the momenta of $\nu_i$ and $\nu_j$ differ by
\beq
p_\nu^i - p_\nu^j = -\frac{1}{2m_\pi} \frac{E_\mu^0}{E_\nu^0} \Delta m^2_{ij}~~.
\label{eq19}
\eeqn
Here, $E_\nu^0$ is the energy the neutrino would have in the pion rest frame if it were massless. From Eq.~(\ref{eq17}) - (\ref{eq19}), it then follows that the propagation phases of two different mass eigenstates $\nu_i$ and $\nu_j$ differ by
\begin{eqnarray}
m_\nu^i \tau_\nu^i - m_\nu^j \tau_\nu^j  &=& [(E_\nu^i - E_\nu^j) - (p_\nu^i -p_\nu^j)]\,L^0 \nonumber \\
 &=&  \frac{\Delta m^2_{ij}}{2m_\pi} \left[ 1 +  \frac{E_\mu^0}{E_\nu^0}\right] L^0 = \Delta m^2_{ij} \frac{L^0}{2E_\nu^0} ~~.
\label{eq20}
\end{eqnarray}
Thus, we may take the neutrino propagation amplitude $\exp (-i m_\nu^i \tau_\nu^i)$ to be
\beq
e^{{\textstyle -i (m_\nu^i)^2 \frac{L^0}{2E_\nu^0} }} ~~,
\label{eq21}
\eeqn
and all the relative phases in our amplitude of \Eq{8} for the process of Fig.~\ref{f2} will be correct.

Using this result, and dropping the muon interaction and propagation amplitudes, which, being $i$-independent, will not affect the neutrino oscillation pattern, we find from \Eq{8} that
\beq
Amp = \sum_{i=1,2,3} U^*_{\mu i} \,e^{{\textstyle -i (m_\nu^i)^2 \frac{L^0}{2E_\nu^0} }} U_{ei} ~~.
\label{eq22}
\eeqn
We recall that the $L^0$ and $E_\nu^0$ appearing in this expression are the neutrino travel distance and energy (neglecting neutrino mass) {\em in the pion rest frame.} However, if neutrino oscillation vs. travel distance {\em in the laboratory frame} is to be observed, then obviously the location in the lab of the neutrino source---the 
pion in the process of Fig.~\ref{f2}---must be known to within an oscillation length or so. But then, by the uncertainty principle $\Delta p \Delta x \ge \hbar$, there must be some spread in lab-frame pion momenta \cite{ref4}. The pions whose decays yield the neutrinos of an oscillation experiment cannot all be at rest. However, because our neutrinos are ultra-relativistic, the travel time of one of them in the pion rest frame is essentially equal to its travel distance $L^0$. Similarly, its momentum in the pion rest frame is equal to its energy $E_\nu^0$. Then, from the Lorentz transformation, the lab-frame travel distance $L$ and energy $E$ of this neutrino are related to their pion-rest-frame counterparts, $L^0$ and $E_\nu^0$, by
\beq
\frac{L}{E} \cong \frac{\gamma_\pi (L^0 + \beta_\pi L^0)}{\gamma_\pi (E_\nu^0 + \beta_\pi E_\nu^0)} = \frac{L^0}{E_\nu^0} ~~.
\label{eq23}
\eeqn
Here, $\beta_\pi$ is the velocity of the parent pion in the lab, and $\gamma_\pi = 1/ \sqrt{1-\beta_\pi^2}$. From \Eq{23}, we see that in terms of lab-frame variables, the Lorentz-invariant amplitude of \Eq{22} is
\beq
Amp = \sum_{i=1,2,3} U^*_{\mu i} \,e^{{\textstyle -i (m_\nu^i)^2 \frac{L}{2E} }} U_{ei} ~~.
\label{eq24}
\eeqn

The sequence of events depicted in Fig.~\ref{f2} is what would usually be called $\nu_\mu \ra \nu_e$ oscillation, with the addition of an interaction between matter and the muon that recoils against the neutrino in the pion decay. We are interested primarily in the probability for the $\nu_\mu \ra \nu_e$ oscillation, integrated over all the possible fates of the muon. Apart from a possible overall normalization factor, this muon-integrated $\nu_\mu \ra \nu_e$ oscillation probability, $P(\nu_\mu \ra \nu_e)$, will be given by the absolute square of the amplitude of \Eq{24}, from which the muon interaction and propagation amplitudes have been removed.
Assuming that the mixing matrix $U$ is unitary, we find from \Eq{24} that
\begin{eqnarray}
P(\nu_\mu \ra \nu_e) & = &   - 4\, \sum_{i>j}  \Re\, (U_{\mu i}^* U_{e i} U_{\mu j} U_{e j}^*)\, \sin^2 (\Delta m^2_{ij} \frac{L}{4E})
	\nonumber   \\
& &+ 2\,\sum_{i>j} \Im \,(U_{\mu i}^* U_{e i} U_{\mu j} U_{e j}^*)\, \sin (\Delta m^2_{ij} \frac{L}{2E}) ~~.
\label{eq25}
\end{eqnarray}
This result agrees completely, even in normalization, with the usual expression for $P(\nu_\mu \ra \nu_e)$ \cite{ref5}. In this paper, we have derived $P(\nu_\mu \ra \nu_e)$ in the same way that we treated $B$-factory experiments. We allowed for the kinematical entanglement between the muon and the neutrino produced in a pion decay \cite{ref6}, but this entanglement proved to be irrelevant. In deriving $P(\nu_\mu \ra \nu_e)$, we did not need to make any assumption about how the energies of the different coherent mass eigenstate components of a neutrino beam are related. Instead, the energies of our coherently contributing mass eigenstates followed from energy-momentum conservation in $\pi \ra \mu \nu$.

\section{Summary} \label{sec4}

In summary, we have treated $B$-factory and neutrino-oscillation experiments, two very interesting examples of the consequences of coherence in quantum mechanics, within a common framework. In the processes studied in $B$-factory experiments, puzzling Einstein-Podolsky-Rosen correlations really do exist, but in our approach one can understand these processes without directly encountering those perplexing correlations. In treating neutrino oscillation, our approach allows for quantum entanglement but yields the standard result for the oscillation probability. It arrives at this result without having to make any non-intuitive assumption about neutrino mass-eigenstate energies.

The approach described in this paper raises some interesting issues that will be explored in future work.

\Acknowledgements

This paper is based in part on work done some time ago with Leo Stodolsky \cite{ref1}, and is an outgrowth of work done recently with Joachim Kopp, Hamish Robertson, and Petr Vogel (B. Kayser \etal\ in Ref.~\cite{ref6}). It is a pleasure to thank all these colleagues for fruitful, illuminating, and enjoyable collaborations, and to thank Leo Stodolsky and Joachim Kopp for very helpful recent discussions. I would also like to thank Abi Soffer, Yossi Nir, and Gilad Perez for the invitations to FPCP 11 and the Weizmann Institute, for hospitality that was generous and warm in many ways, and for gracious patience. I am indebted to Susan Kayser for expert technical support. Fermilab is operated by Fermi Research Alliance, LLC under Contract No. DE-AC02-07CH11359 with the US Department of Energy.

\end{document}